\documentclass[twocolumn,tighten,times,twocolappendix]{aastex631}
\newcommand{\Halpha}{{H$\alpha$}}

\newcommand{\Lyalpha}{{Ly$\alpha$}}
\newcommand{\Lybeta}{{Ly$\beta$}}
\newcommand{\FeI}{{Fe I}}

\newcommand{\soutdis}[1]{}

\received{October 23, 2023}
\accepted{November 17, 2023}

\shorttitle{Energizing small-scale loops through surface convection}
\shortauthors{N\'obrega-Siverio et al.}

\begin{document}

\title{\Large{Deciphering the solar coronal heating:\\ 
Energizing small-scale loops through surface convection}}

\correspondingauthor{D. N\'obrega-Siverio}
\email{dnobrega@iac.es}

\author[0000-0002-7788-6482]{D. N\'obrega-Siverio}
\affiliation{Instituto de Astrof\'isica de Canarias, 
            E-38205 La Laguna, Tenerife, Spain}
\affiliation{Universidad de La Laguna, Dept. Astrof\'isica, 
            E-38206 La Laguna, Tenerife, Spain}
\affiliation{Rosseland Centre for Solar Physics, University of Oslo, 
            PO Box 1029 Blindern, 0315 Oslo, Norway}
\affiliation{Institute of Theoretical Astrophysics, University of Oslo, 
            PO Box 1029 Blindern, 0315 Oslo, Norway}

\author{F. Moreno-Insertis}
\affiliation{Instituto de Astrof\'isica de Canarias, 
            E-38205 La Laguna, Tenerife, Spain}
\affiliation{Universidad de La Laguna, Dept. Astrof\'isica, 
            E-38206 La Laguna, Tenerife, Spain}

\author[0000-0001-8882-1708]{K. Galsgaard}
\affiliation{School of Mathematics and Statistics, University of St Andrews, 
             St Andrews KY16 9SS, Scotland, UK}

\author[0000-0002-0922-7864]{K. Krikova}
\affiliation{Rosseland Centre for Solar Physics, University of Oslo, 
             PO Box 1029 Blindern, 0315 Oslo, Norway}
\affiliation{Institute of Theoretical Astrophysics, University of Oslo, 
            PO Box 1029 Blindern, 0315 Oslo, Norway}

\author[0000-0003-2088-028X]{L. Rouppe van der Voort}
\affiliation{Rosseland Centre for Solar Physics, University of Oslo, 
             PO Box 1029 Blindern, 0315 Oslo, Norway}
\affiliation{Institute of Theoretical Astrophysics, University of Oslo, 
            PO Box 1029 Blindern, 0315 Oslo, Norway}

\author[0000-0003-0020-5754]{R. Joshi}
\affiliation{Rosseland Centre for Solar Physics, University of Oslo, 
             PO Box 1029 Blindern, 0315 Oslo, Norway}
\affiliation{Institute of Theoretical Astrophysics, University of Oslo, 
            PO Box 1029 Blindern, 0315 Oslo, Norway}

\author[0000-0001-9806-2485]{M. S. Madjarska}
\affiliation{Max Planck Institute for Solar System Research, 
             Justus-von-Liebig-Weg 3, 37077 Göttingen, Germany}
\affiliation{Space Research and Technology Institute, Bulgarian Academy of Sciences,
             Acad. Georgy Bonchev Str., Bl. 1, 1113, Sofia, Bulgaria}

\begin{abstract}
The solar atmosphere is filled with clusters of hot small-scale loops commonly known as 
Coronal Bright Points (CBPs).
These ubiquitous structures stand out in the Sun 
by their strong X-ray and/or extreme-ultraviolet (EUV) emission for hours to days,
which makes them a crucial piece when solving the solar coronal heating puzzle.
In addition, they can be the source of coronal jets and small-scale filament eruptions.
Here we present a novel 3D numerical model using the Bifrost code 
that explains the sustained CBP heating for several hours. 
We find that stochastic photospheric convective motions alone significantly stress the CBP 
magnetic field topology, leading to important Joule and viscous heating concentrated around the CBP's inner 
spine at a few megameters above the solar surface.
We also detect continuous upflows with faint EUV signal resembling observational dark coronal jets and 
small-scale eruptions when \Halpha\ fibrils interact with the reconnection site.
We validate our model by comparing simultaneous CBP observations from SDO and SST with 
observable diagnostics calculated from the numerical results for EUV wavelengths
as well as for the \Halpha\ line using the Multi3D synthesis code.
Additionally, we provide synthetic observables to be compared with Hinode, Solar Orbiter, and IRIS.
Our results constitute a step forward in the understanding of the many different facets of the solar coronal 
heating problem.
\end{abstract}


\section{Introduction}\label{sec:intro}
Understanding why the solar corona temperature is a few hundred times larger than that at the surface requires 
studying the heating mechanisms across different solar regions and phenomena 
\citep[e.g.,][]{Parker:1972,Galsgaard_Nordlund:1996,Gudiksen_Nordlund:2005,Klimchuk:2006,Parnell_De-Moortel:2012}.
A particularly important case is that of the Coronal Bright Points or CBPs,
structures with projected sizes between 5 and 40~Mm that stand out with enhanced EUV and X-ray
emission against the quiet Sun \citep{Golub_etal:1974,Madjarska:2019}.
CBPs are often observed to appear above 
photospheric regions with a strong parasitic polarity surrounded by a network predominantly of the opposite polarity, a situation that typically leads to a fan-spine magnetic structure with a nullpoint at several megameters in the corona \citep{Zhang_etal:2012, Mou_etal:2016,Galsgaard_etal:2017,Madjarska_etal:2021,Cheng_etal:2023}.
Their relevance stems from their copious energy emission, representing the main contributor to high-energy radiation 
over the solar disk outside active regions \citep{Mondal_etal:2023}; as well as their nearly uniform
presence in the Sun and their long lifetime, from hours to days \citep{Madjarska:2019}.
Considered to be downscaled versions of active regions \citep{Madjarska:2019,Gao_etal:2022}, CBPs are composed of 
small-scale loops, making them a relevant case study for this kind of fundamental magnetic field topology 
\citep{Rappazzo_etal:2008,Reale:2014}. 
Moreover, CBPs can be the source regions of eruptions of both hot and cool plasma 
\citep{Hong_etal:2014,Sterling_etal:2015,Kumar_etal:2019,Madjarska_etal:2022},
which may constitute a significant  input of energy and mass for the solar corona 
\citep{Lionello_etal:2016,Viall_Borovsky:2020}.  
%

A significant theoretical effort has been devoted to studying the heating of CBPs through (a) analytical studies 
collectively known as Converging Flux Models \citep{Priest_etal:1994,Dreher_etal:1997,Priest_etal:2018}; (b) one-dimensional models 
of individual magnetic loops \citep{Reale:2014}; or (c) by means of multi-dimensional purely magnetohydrodynamical 
(MHD) models, mainly analyzing the energy conversion during magnetic reconnection \citep{Galsgaard_etal:2000,
von-Rekowski_etal:2006a,Santos_Buchner:2007,Javadi_etal:2011,Wyper_etal:2018b,Syntelis_etal:2019}.
%
%
%
A drawback of these approaches is that they rely on ad-hoc driving mechanisms,
imposing converging flows or large-scale surface motions in order to drive magnetic reconnection at the coronal nullpoint and get a CBP.
These mechanisms do not reflect the actual 
stochastic convective flows in the solar photosphere, which have been shown 
to provide enough energy to heat the corona in previous numerical simulations \citep[e.g.,][]{Gudiksen_Nordlund:2005,Rempel:2017,Chen_etal:2022}.
However, as we propose in this paper, the energy injection through surface convection may indeed be the source of the CBP heating.
An additional drawback of previous CBP models is that they did not include
an appropriate radiation treatment, thus ignoring important entropy sources in the lower atmosphere.
This not only precludes the calculation of many basic observational diagnostics but also restricts the possibility 
of understanding the heating and cooling processes in the chromosphere.

In this Letter, we present a comprehensive 3D radiative-MHD model of a CBP based on a magnetic nullpoint 
configuration. 
%
Our model shows that a CBP can be powered for at least several hours through the 
continuous action of the underlying surface convection. 
The inclusion of the relevant radiation and entropy sources in our model allows us
to analyze the associated deposition of energy in different layers of the solar atmosphere and the corresponding changes 
in the dynamics, including the study of hot ejections and small-scale eruptive chromospheric phenomena
within the CBP.
To validate our theoretical model, we also present simultaneous observations of a representative CBP from space (SDO)
and from the ground (SST).
%
Through forward modeling, 
we successfully reproduce the main observed
features from SDO and SST, confirming the adequacy of our model across
different atmospheric layers.

\section{Results} \label{sec:results}

\subsection{Overview of the Experiment}\label{sec:overview}
\begin{figure*}[!ht]
        \centering
        \includegraphics[width=1.00\textwidth]{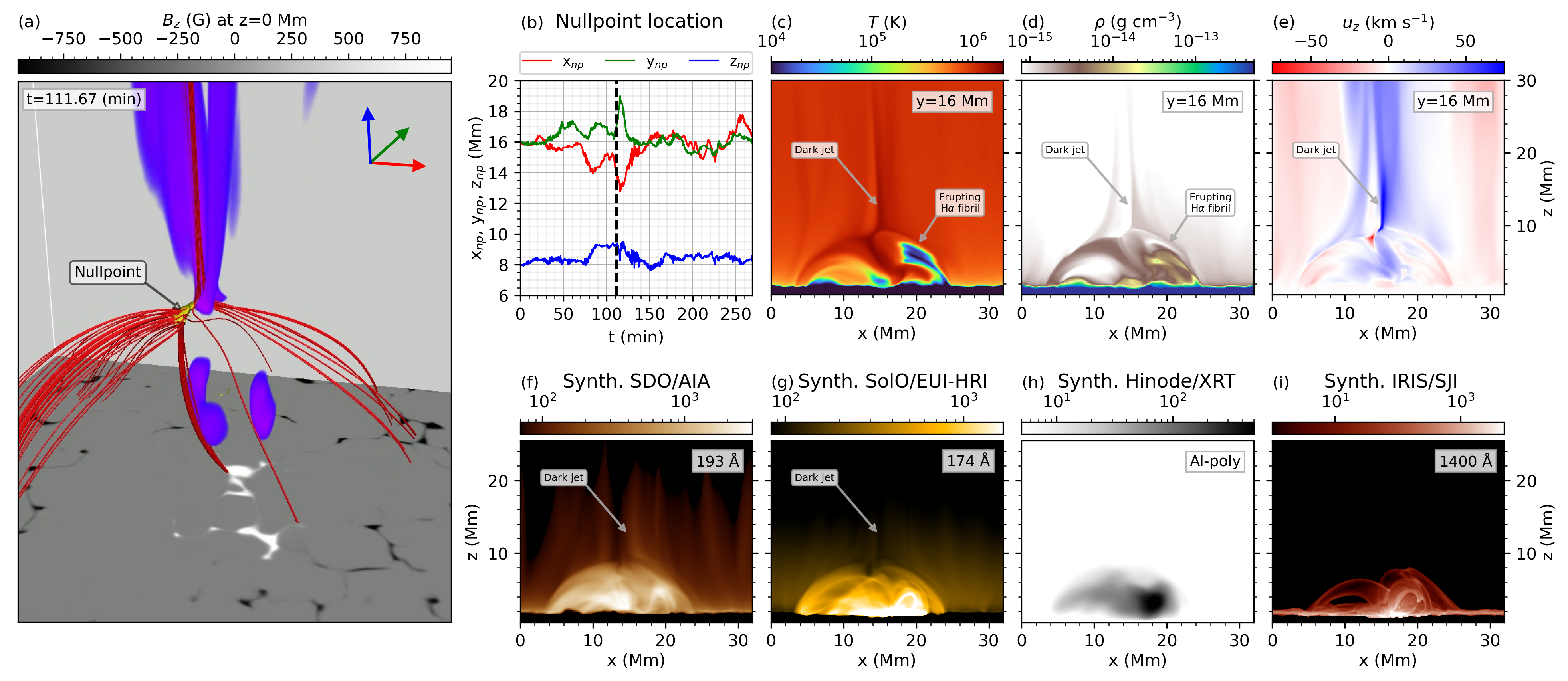}
        \caption{Experiment overview.
        (a)
        3D magnetic nullpoint topology
        (red lines) at $t=111.67$~min superimposed on a horizontal 
        map of $B_z$ at $z=0$.
        Blue-purple volume rendering corresponds to upflows with velocities from 45-75 km s$^{-1}$.
        The yellow isosurface at $B=1$ G harbors the nullpoint location at the center of the image.
        The red-green-blue coordinate system indicates the x-y-z axis orientation.
        (b) Time-evolution of the nullpoint coordinates (x$_{np}$, y$_{np}$, z$_{np}$).
        (c), (d), (e) Temperature, density, and vertical velocity cuts at y$=$16~Mm, respectively.
        (f), (g), (h), (i) Synthetic response integrated along the y-axis to mimic a limb observation 
        by SDO/AIA~193~\AA, SolO/EUI-HRI~174~\AA, Hinode/XRT Al-poly (reversed color scale), and IRIS/SJI~1400~\AA, 
        respectively, in DN s$^{-1}$ pix$^{-1}$ units.
        The associated animation comprises the whole experiment evolution from $t=0$ to $t=268.33$~min. \\
        (An animation of this figure is available.)
        } 
        \label{fig:01}
\end{figure*}

The initial configuration consists of a rotationally-symmetric 3D 
null point magnetic topology (see Figure~\ref{sup:01} in Appendix \ref{app:numerical_experiment}, where details concerning the initial condition and the numerical code are also given). 
The inner spine of the structure is rooted in a region with a parasitic positive polarity, while the fan structure 
is anchored in the surrounding field of the opposite sign. 
A general outlook of the subsequent evolution of the experiment can be gained through Figure~\ref{fig:01} and its associated animation.
The initial topology structure is rapidly disrupted by the stochastic granulation in the photosphere leading to a 
magnetic network pattern of the predominant polarity at the surface; the overall magnetic field topology above is
stressed and magnetic reconnection is triggered at the nullpoint.
Simultaneously, the magnetic loops underneath the fan surface experience intense heating, giving rise to a 
CBP with conspicuously enhanced emissivity in EUV, X-rays, and UV, as evidenced in Figure~\ref{fig:01} through the synthetic observables of panels (f), (g), (h), and (i) (see Appendix~\ref{app:synthesis} for the calculation details).
The diameter of the CBP, estimated from the EUV response at its base, is approximately 20~Mm, which is within the typical range reported in observations \citep[5--40~Mm, see][and references therein]{Madjarska:2019}.

The CBP nullpoint undergoes continuous perturbations, as illustrated in panel~(b) through the time-evolution 
of its coordinates (x$_{np}$, y$_{np}$, z$_{np}$), which can be shifted by up to a few Mm, similarly to the nullpoint 
displacements inferred from observations \citep{Galsgaard_etal:2017}.
Around $t=111.67$~min, the time of the snapshots in Figure~\ref{fig:01}, the current sheet around the nullpoint 
experiences a major disturbance, concerning both its location and shape.
The reason for this perturbation is the approach of an elongated cool and dense structure to the reconnection site 
visible in  panels (c) and (d). 
This structure is shown to be an \Halpha\ chromospheric fibril in Section~\ref{sec:synthesis}.
Throughout the evolution of the CBP, other chromospheric fibrils and spicule-like structures are found to develop 
in the lower atmosphere (see animation at $t=47,\ 107,\ 153,\ 221$~min), reaching the reconnection 
site and affecting the CBP dynamics as well as its brightness.
Such features could be the result of the interaction of chromospheric waves and shocks with the CBP magnetic structure. 
A detailed analysis will be covered in a follow-up paper.

Rapid upflows can be seen associated with the CBP all along the experiment, as visualized through the blue-purple 
volume rendering in panel~(a), 
which show velocities from 45 to 75 km s$^{-1}$, and through the vertical velocity $u_z$ in panel~(e).
These upflows are mainly concentrated around the spines of the CBP; those along the outer spine are basically reconnection outflows emanating from the nullpoint, whereas the ones around the inner spine are probably due to channeling of magnetosonic waves.
The flows arising from the CBP nullpoint, like the one at the time
of the figure, are of particular interest for the corona.
%
These hot (1--2 MK) collimated ejections have low contrast in density compared to the surrounding plasma, and 
therefore they are not (or barely) distinguishable in the synthetic EUV images 
(see discussion in Section~\ref{sec:discussion}).

\subsection{Energization}\label{sec:energization}
\begin{figure*}[!ht]
        \centering
        \includegraphics[width=1.00\textwidth]{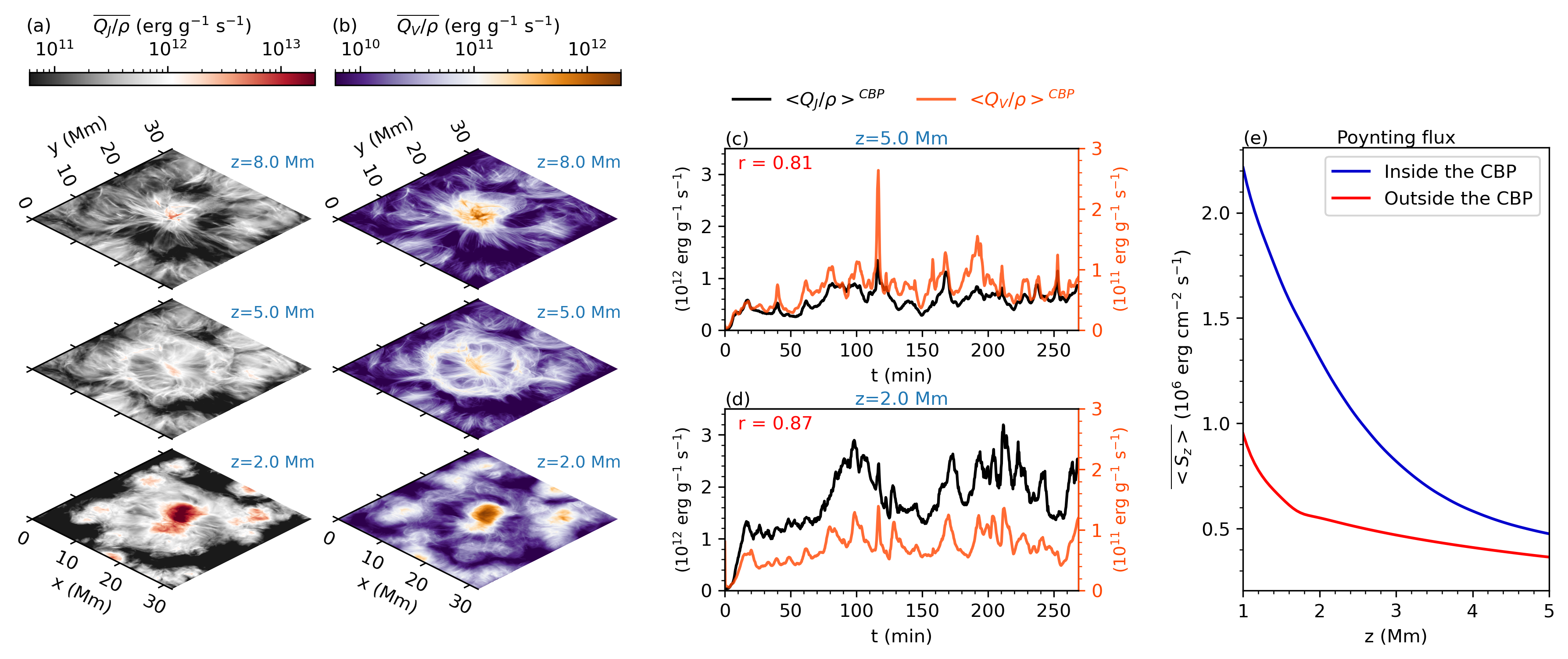}
        \caption{Heating in our CBP model.
        (a), (b) Heating per unit mass averaged over $\sim4.5$ hours 
        for the Joule ( $\overline{\!Q_{J}/\rho\!}$ ) and viscous 
        ( $\overline{\!Q_{V}/\rho\!}$ ) terms, respectively, at $z = [8, 5, 2]$~Mm.  
        (c), (d) Heating per unit mass averaged horizontally within the CBP at $z = [5, 2]$~Mm, respectively, 
        for the Joule ($<\!Q_{J}/\rho\!>^{\mathrm{CBP}}$, black) and
        viscous ($<\!Q_{V}/\rho\!>^{\mathrm{CBP}}$, orange) terms as a function of time.
        The Pearson correlation coefficient $r$ between the curves is shown in red in the upper 
        left corner of the plot frame.
        (e) Time average of the horizontal mean vertical Poynting flux ($\overline{<\!S_{z}\!>}$) 
        within the CBP (blue) and in the coronal hole surroundings (red).
        } 
        \label{fig:02}
\end{figure*}

To address the fundamental question about the sustained CBP heating and its location, we have first analyzed the 
heating per unit mass due to the Joule and viscous terms.
Panels (a) and (b) of Figure~\ref{fig:02} contain time-averaged results over approximately 4.5 hours of the simulation at three different heights, $z = [8, 5, 2]$~Mm, thus encompassing the regions in close 
proximity to the CBP nullpoint downwards to the low atmosphere.
Both the Joule heating (left column) and viscous heating (right column) primarily occur at lower heights, at a few megameters above the solar surface, concentrated 
around the inner spine, in the region dominated by the strong parasitic positive polarity.
We found that in these regions the electric current vector subtends a small angle with respect to the 
concentrated nearly vertical magnetic field, indicating that the magnetic configuration is close to force-free there.
Heating is also found, albeit to a lesser extent, in the magnetic patches outside of the fan surface in the 
surroundings of the CBP at $z=2$~Mm (bottom panels), as well as in the region near the nullpoint at $z=8$~Mm 
(top panels). 
Thus, our findings indicate that CBP loops are primarily heated in the low atmosphere, with a secondary contribution 
at coronal heights near the reconnection site. 
Furthermore, there is a remarkable spatial correlation between the Joule and viscous heating across various heights. 
This suggests that regions exhibiting the highest electric current intensity and velocity gradient typically coincide.
As a consequence, our results concerning the location of the CBP heating could be general, regardless of which mechanism (Joule or viscous) is dominant, even if in our experiment the Joule heating term is approximately one order of magnitude greater than the viscous one (see discussion in Section~\ref{sec:discussion}).

Another important question for CBPs, and for the corona in general, concerns whether the heating occurs on short 
or long timescales \citep{Reale:2014}.
To pursue this question, panels (c) and (d) of Figure~\ref{fig:02} illustrate, as a function of time, the horizontal 
average within the CBP fan surface of the Joule (black) and viscous (orange) 
heating per unit mass at $z=[2, 5]$~Mm, respectively.
Both heating mechanisms seem to be continuously at work, showing fluctuations with different time scales.
For instance, the pronounced peak observed mainly in the viscous heating profile at $z=5$~Mm at 
$t \in [110,120]$~min corresponds to the \Halpha\ fibril eruption explained in the previous section.
In general, there is a clear time correlation between the Joule and viscous dissipation mechanisms. 
In fact, the Pearson correlation coefficient $r$ between them is greater than $0.8$ for the different 
heights we have analyzed.

As a final point, we consider the energy injection through the Poynting flux. 
Panel~(e) of Figure~\ref{fig:02} contains the mean vertical Poynting flux computed within the CBP fan surface and 
averaged in time, $\overline{<\!S_{z}\!>}$,  as a function of height (blue curve).
This plot strongly suggests that the heating of the plasma within the CBP domain is due to the deposition of the 
upward Poynting flux, which, itself, results from the stochastic photospheric motions stressing the CBP magnetic
field. 
This implies that no major organized photospheric flows (leading, e.g., to spine dragging or convergence of 
magnetic polarities) are necessary to maintain the heating of the CBPs.
Importantly, the average $\overline{<\!S_{z}\!>}$ within the CBP is larger at all the heights than the corresponding 
quantity in the surroundings (red curve), and the same applies to the corresponding gradient in height 
(in absolute value).
%
The greater vertical Poynting flux obtained within the CBP is due to the strong magnetic concentration in the parasitic polarity.
The Poynting flux values and the height gradient are in agreement with those indicated by other
radiative-MHD numerical experiments as necessary to sustain heating in the quiet Sun corona 
\citep{Rempel:2017,Chen_etal:2022}.
In addition, these values seem to fulfill the energy requirements for a hot corona inferred 
from observations 
\citep[e.g., $0.8\times10^6$ erg cm$^{-2}$ s$^{-1}$ for coronal holes,][]{Withbroe_Noyes:1977}.
Moreover, computing the surface integral of the vertical Poynting flux within the fan surface domain 
over two horizontal planes in the corona (e.g., $z=2$~Mm and $z=5$~Mm), 
we obtain a magnetic energy deposition in the CBP of 
$\approx7\times10^{24}$ erg s$^{-1}$,
which is more than enough to explain the rough estimate of the CBP energy losses obtained through observations 
 \citep[$10^{23}$ to $10^{24}$~erg s$^{-1}$,][]{Habbal_Withbroe:1981}.
The loss of electromagnetic energy through the fan surface is most probably small by comparison, at any rate when 
averaging in time.

\subsection{Confronting Observations and Modeling}\label{sec:synthesis}
\begin{figure*}[!ht]
        \centering
        \includegraphics[width=1.00\textwidth]{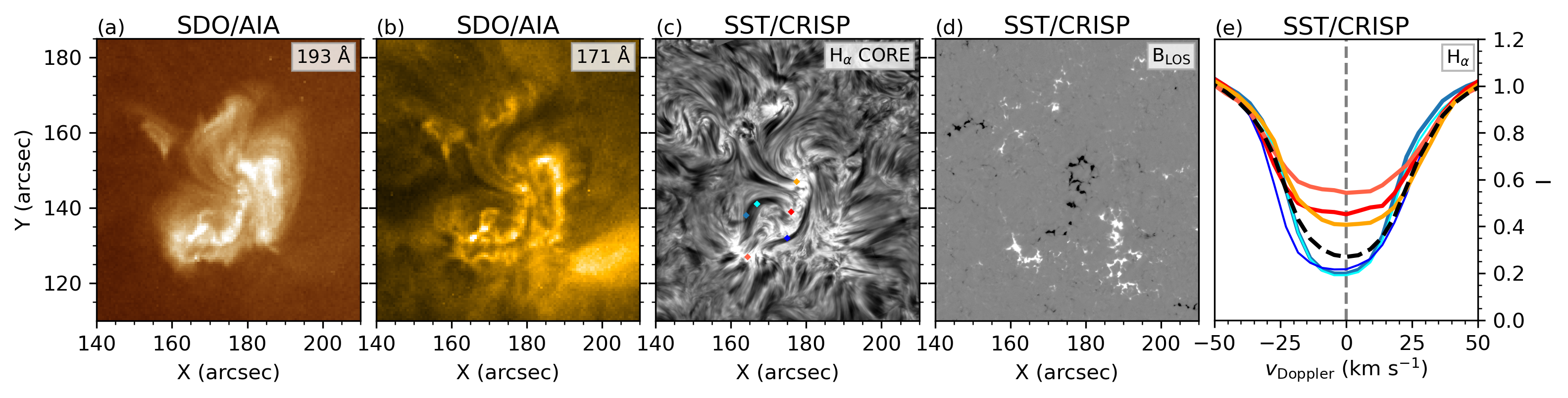}
        \caption{Observations of a representative CBP on 2022 July 01 at 08:08:52 UT.
        (a), (b) Hot coronal response in EUV detected by SDO/AIA~193~\AA\ and SDO/AIA~171~\AA, respectively.
        (c) Cool chromospheric structure of the CBP in the \Halpha\ line core observed with SST/CRISP. 
        (d) Photospheric line-of-sight (LOS) magnetic field (B$_{\mathrm{LOS}}$) from SST/CRISP.
        (e) \Halpha\ profiles for different regions of the dark fibrils (cool colors) and core  
        brightenings (warm colors) marked in panel~(c). 
        The average profile over the whole field-of-view is shown as a black dashed line.
        The intensity of the profiles has been normalized to the intensity value at $50$~km~s$^{-1}$.
        } 
        \label{fig:03}
\end{figure*}

To reinforce the previous results, we present here SDO and SST observations (see Appendix~\ref{app:observations}) of 
a CBP that occurred on 2022 July 01 located at heliocentric coordinates $(x,y) = (177\arcsec,147\arcsec)$. 
The million-K coronal response of the CBP, shown in the SDO/AIA images (panels (a) and (b) of 
Figure~\ref{fig:03}), has a compact and bright rowel-like shape that covers a horizontal domain of $\sim40\arcsec\times40\arcsec$.
The photospheric line-of-sight magnetic field image from SST/CRISP in panel~(d) shows a 
parasitic negative polarity at the center of the CBP, surrounded by a dominant opposite polarity magnetic region, a
configuration that can naturally lead to a fan-spine configuration with a nullpoint at coronal heights
\citep{Zhang_etal:2012,Mou_etal:2016,Galsgaard_etal:2017,Madjarska_etal:2021,Cheng_etal:2023}.
The chromospheric \Halpha\ core image from SST/CRISP in panel~(c) indicates that the CBP contains a bunch 
of dark and long \Halpha\ fibrils that connect opposite polarities, analogous to fibrils reported in lower resolution 
observations \citep{Madjarska_etal:2021}; and there are strong brightenings mainly located at the base of the 
fibrils, possibly indicative of chromospheric heating.
The spectral profiles at different locations of the fibrils and brightenings are shown in panel~(e).

\begin{figure*}[!ht]
        \centering        
        \includegraphics[width=1.00\textwidth]{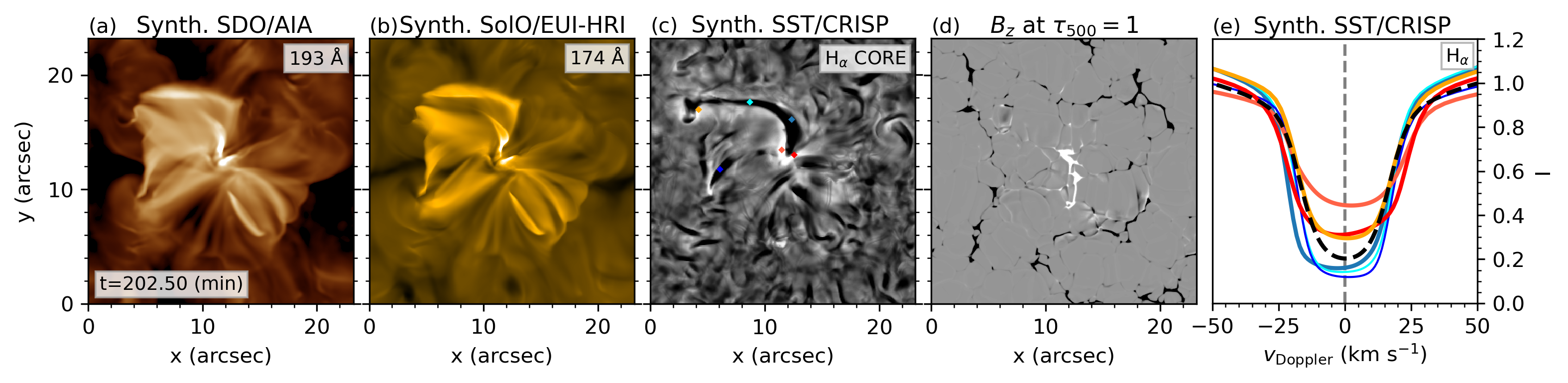}
        \caption{Forward modeling from our 3D CBP numerical experiment as observed on-disk.
        (a), (b) Synthetic EUV images for SDO/AIA~193~\AA\ and SolO/EUI-HRI~174~\AA, respectively.
        (c) Synthetic \Halpha\ response in the core of the line as observed by SST/CRISP.
        (d) Vertical magnetic field $B_z$ at the solar surface 
        ($\tau_{500\thinspace{\rm nm}}=1$). 
        (e) \Halpha\ profiles for different regions of the dark fibrils (cool colors) and core  
        brightenings (warm colors) marked in panel~(c). 
        The average profile over the whole box is shown as a black dashed line.
        The intensity of the profiles has been normalized to the intensity value at $50$~km~s$^{-1}$.
        The associated animation shows the synthetic results for $t=111.67$, $202.50$, and $264.17$~min.\\
        (An animation of this figure is available.)
        } 
        \label{fig:04}
\end{figure*}

In order to prove that our model satisfactorily reproduces the observations, Figure~\ref{fig:04} and its associated
animation contain the forward-modeling results from the simulations.
Panel~(a) shows that we indeed obtain a rowel-like structure composed of hot loops with enhanced EUV response 
similar to the SDO/AIA 193~\AA\ observations, although our CBP is smaller than the observed one. 
Given its current interest, in panel~(b) we also add the EUV response as would be observed by Solar Orbiter's 
EUI instrument, with its higher spatial resolution. 
The comparison to observations is also successful for the chromospheric counterpart, as seen in panel~(c).
Our CBP model is realistic enough to reproduce a pattern of elongated dark fibrils, as observed in the
\Halpha\ images; and the existence of strong brightenings 
at the base of these fibrils, surely a manifestation of the location of the heating in the lower atmosphere, 
as discussed in the previous section. 
The agreement is further illustrated by comparing panels (e) of Figures~\ref{fig:03} and \ref{fig:04}. 
The observational and synthetic spectral profiles show a similar behaviour regarding the \Halpha\ line core 
intensity at different locations of the fibrils and brightenings, namely, the fibrils are characterized by a 
less bright \Halpha\ core than the average with a small Doppler shift, while the brightenings show an enhanced 
\Halpha\ core.
The smaller width in the synthetic profiles may be due to the modest numerical resolution not capturing 
the highly dynamic complexity of the chromosphere \citep{Hansteen_etal:2023}.
At the photosphere, the vertical magnetic field of the model (panel~(d) of Figure~\ref{fig:04}) shows that the 
elongated \Halpha\ fibrils connect the central parasitic polarity with the strong, opposite-polarity magnetic 
patches surrounding the CBP, as shown in our observations (panel~(d) of Figure~\ref{fig:03}) 
as well as in other CBP observations \citep{Madjarska_etal:2021}.

\section{Discussion} \label{sec:discussion}

In this Letter, through a 3D radiative-MHD model using the Bifrost code, we have shown that CBPs can be energized 
by the action of the continuous surface convection motions, thus obviating the need for major, 
organized photospheric flows such as converging flows \citep{Priest_etal:1994,Dreher_etal:1997,Priest_etal:2018,Syntelis_etal:2019}
or large-scale surface flows \citep{Wyper_etal:2018b}.
Our investigation reveals that the CBP loops are heated by both Joule and viscous heating predominantly 
in the lower atmosphere,  with a secondary contribution at coronal heights close to the reconnection site. 
This is a major difference with 2D models \citep[e.g.,][]{Syntelis_etal:2019,Nobrega-Siverio_Moreno-Insertis:2022}, 
which can only capture the heating at the reconnection site since mechanisms, such as 
magnetic fieldline braiding, are intrinsically three-dimensional.
The heating rate from both the resistive and viscous heating terms is intermittent, which seems to be consistent 
with findings from simulations focused on individual coronal loops \citep{Breu_etal:2022}.
Furthermore, both the spatial and time correlation found between the Joule and viscous heating within the CBP suggests 
that the heating results obtained in this Letter may be widely independent of the value of the magnetic 
Prandtl number, in line with the conclusions reached in recent coronal numerical simulations 
\citep{Rempel:2017,Chen_etal:2022}.
This implies that, even though current numerical models are far from reproducing the Prandtl numbers of the solar atmosphere, our results may well be applicable to understand the heating of small-scale loops (CBPs) and their properties.

By means of X-ray, EUV, and UV forward modeling, we have shown that the CBP's multi-wavelength response and 
corresponding brightness exhibit fluctuations on the scale of minutes (see animation of Figure~\ref{fig:01}), 
resembling intensity variations reported in observations \citep{Ugarte-Urra_etal:2004a,Doschek_etal:2010,
Kumar_etal:2011,Ning_Guo:2014,Zhang_etal:2014,Chandrashekhar_Sarkar:2015,Kayshap_Dwivedi:2017,Gao_etal:2022}.
Our results also indicate that chromospheric phenomena, such as \Halpha\ fibrils and spicule-like features, 
can perturb the CBP, leading to significant changes in the EUV intensity.
These findings support the recent results about the impact of spicules in 
the CBP brightness variations \citep{Nobrega-Siverio_Moreno-Insertis:2022,Bose_etal:2023}.
Additionally, our model demonstrates that CBPs can produce collimated hot ejections that do not show 
appreciable emission in EUV due to the slight density contrast with the surroundings. 
This may offer theoretical support to the observed features rooted in CBPs named \textit{dark jets} that have 
been detected in EUV spectroscopic data using Hinode/EIS, but with either weak or absent signatures in 
SDO/AIA 193~\AA\ images \citep{Young:2015}.
This may be also the key to explaining the lack of clear observable jetting activity in EUV 
images in the early stages of CBPs \citep{Kumar_etal:2019}.
Further observational evidence using off-limb examples and spectroscopic observations is needed.
In this vein, our results could provide valuable insights for the near future missions 
Multi-slit Solar Explorer \citep[MUSE,][]{De-Pontieu_etal:2022,Cheung_etal:2022} and Solar-C/EUVST 
\citep[][]{Shimizu_etal:2020}.

In this Letter, we have also provided simultaneous observations of a representative CBP from SDO and SST.
Thus, we not only address the scarcity of observational results about the chromosphere below CBPs 
\citep{Madjarska_etal:2021,Bose_etal:2023}, but also we impose a case-study to test our model.
The similarities between the synthetic observables from the model and the observations are striking, even 
reproducing chromospheric features such as the brightenings in the core of the \Halpha\ line, the dark and elongated
\Halpha\ fibrils, as well as their magnetic photospheric connections. 
By increasing the numerical resolution, we expect to achieve even greater agreement with the observations in aspects such as the width of the \Halpha\ line and the intricate fine structure within CBPs.
All the results described above are a good indication that we are including 
the most important physical ingredients to explain the evolution and heating of CBPs as well as reproducing
their main identifying signatures.

\hfill \break 
\indent
This research has been supported by the European Research Council through the
Synergy Grant number 810218 (``The Whole Sun'', ERC-2018-SyG);
by the Spanish Ministry of Science, Innovation and Universities through project PGC2018-095832-B-I00; 
and by the Research Council of Norway (RCN) through its Centres of Excellence scheme, project number 262622.
The authors acknowledge the computer resources at the MareNostrum
supercomputing installation and the technical support provided by the
Barcelona Supercomputing Center (BSC, RES-AECT-2021-1-0023,
RES-AECT-2022-2-0002). 
The use of UCAR's VAPOR software \citep{VAPOR_1,VAPOR_2} is gratefully acknowledged. 
This work also benefited from discussions at the International Space Science Institute (ISSI) in Bern, 
through ISSI International Team project \#535 \textit{Unraveling surges: a joint perspective from numerical models, 
observations, and machine learning}.
L.R.v.d.V. is supported by RCN project number 325491. 
M.M. acknowledges financial support by DFG grant WI 3211/8-1. 
The Swedish 1-m Solar Telescope is operated on the island of La
Palma by the Institute for Solar Physics of Stockholm University in the Spanish
Observatorio del Roque de Los Muchachos of the Instituto de Astrof\'isica
de Canarias. The Institute for Solar Physics is supported by a grant for research
infrastructures of national importance from the Swedish Research Council (registration
number 2017-00625). 
SDO observations are courtesy of NASA/SDO and the AIA, EVE, and HMI science teams.
The authors thank Dr. Frédéric Auchère for his help to compute the synthetic observables for SolO/EUI-HRI.

\appendix

\section{NUMERICAL EXPERIMENT}\label{app:numerical_experiment}
\restartappendixnumbering

\subsection{Code}
The experiment was carried out using the radiation-magnetohydrodynamics Bifrost code \citep{Gudiksen_etal:2011}.
This code includes, in a self-consistent manner, radiative transfer from the photosphere to the corona; approximate 
recipes for the main radiative losses in the chromosphere due to neutral hydrogen, singly-ionized calcium, and 
singly-ionized magnesium; field aligned thermal conduction, which tends to an
isotropic conductivity prescription for regions with weak magnetic field such as nullpoints; optically thin losses; and an equation of state considering the $16$ most relevant atomic elements for the solar atmosphere either because of their high abundance or because of their importance as electron donors at low temperatures. 
Explicit resistivity and viscosity are included in Bifrost through hyper-diffusion terms \citep{Gudiksen_etal:2011}, following the idea originally described by \cite{Nordlund_Galsgaard:1995}.
A comprehensive description of the resistive and viscous hyper-diffusion operator and the associated free parameters can be found in the papers by \cite{Faerder_etal:2023a,Faerder_etal:2023b}, including a comparative analysis of the hyper-diffusive resistivity versus other resistivity models.

\subsection{Initial Condition}
\begin{figure}[ht]
        \centering
        \includegraphics[width=0.50\textwidth]{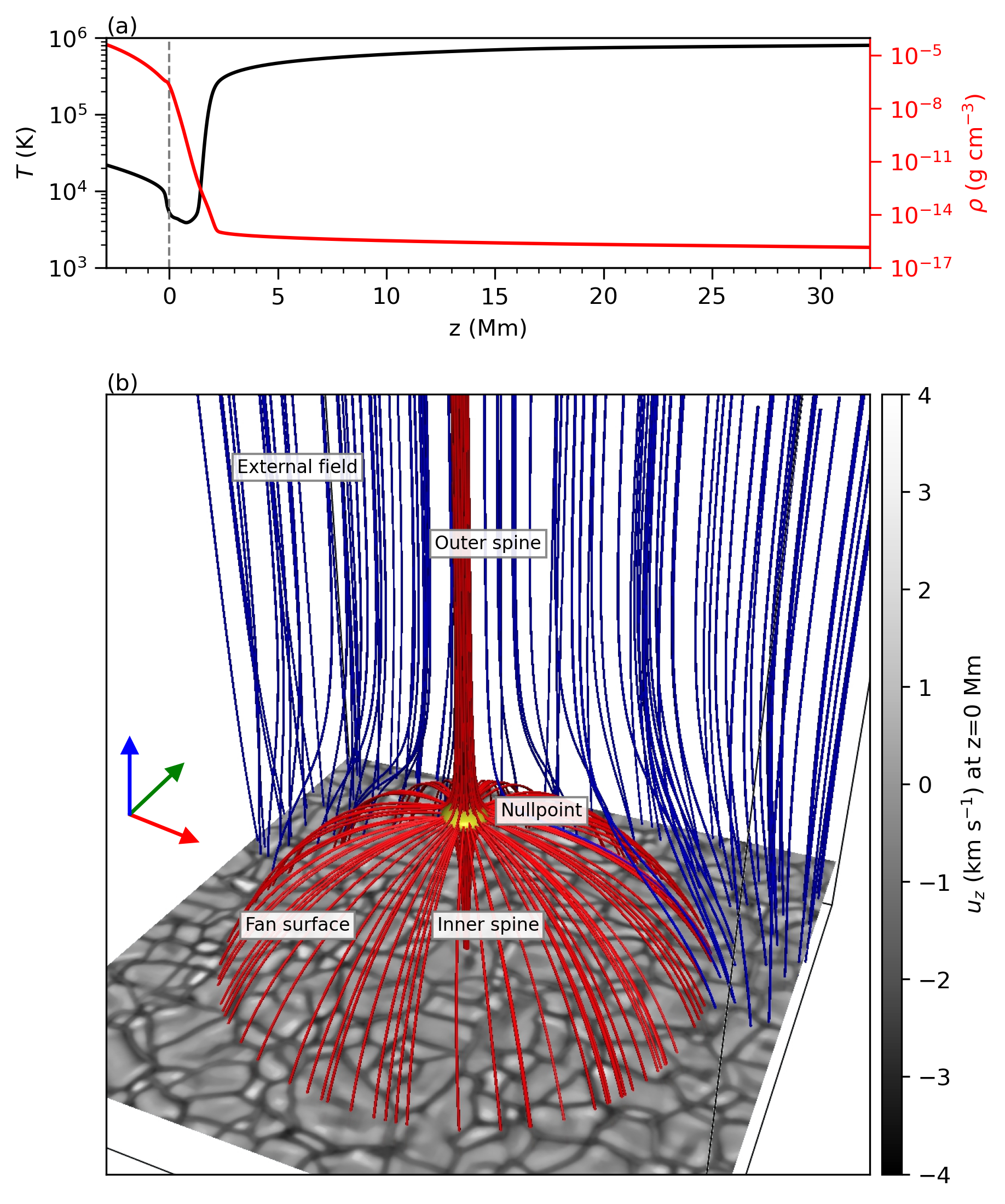}
        \caption{Initial condition.
        (a) Initial horizontal average stratification for the temperature $T$  
        and density $\rho$.
        (b) Imposed nullpoint magnetic field configuration (red lines)
        embedded within a coronal-hole like magnetic field (blue).
        The yellow isosurface at $B=1$~G delimits the nullpoint location.
        Other features of this configuration are also highlighted following
        the 3D nullpoint terminology \citep{Priest_Titov:1996}. 
        The solar granulation pattern of the initial model is 
        shown through the vertical velocity $u_z$ at $z=0$.
        The red-green-blue coordinate system indicates the orientation of the x-y-z axis.
        } 
        \label{sup:01}
\end{figure}

The initial condition was built using a weakly magnetized (0.2~G at the corona) preexisting 3D numerical experiment 
that had reached a statistically stationary equilibrium.
In the vertical direction, the numerical box covers a region from the uppermost layers of the solar interior to 
the corona, that is, $-2.9$~Mm $\leq z\leq 32.3$~Mm, with $z=0$ corresponding to the solar surface (more precisely 
the horizontal surface where $<\hskip -2pt \tau_{500\thinspace{\rm nm}}\hskip -2pt>$=1).
The horizontal extent is $32 \times 32.0$~Mm$^2$, with both $x$ and $y$ starting at the origin of coordinates $(0,0)$.
The domain is resolved with $512\times512\times512$ cells, employing a uniform grid in the horizontal directions 
with spacing $\Delta x= \Delta y= 62.5$~km; and a non-uniform mesh for $z$, using $\Delta z = 50$~km from the interior up
to 12 Mm in the corona, and then progressively decreasing the resolution down to $\Delta z =150$~km at the top of the corona.
The boundary conditions are periodic in the horizontal direction.
In the vertical direction, at the top, characteristic boundary conditions are applied; at the bottom, constant 
entropy is set for the incoming plasma to keep solar-like convection, while the rest of the variables are 
extrapolated.
Panel~(a) of Figure~\ref{sup:01} shows the horizontal averages of the temperature $T$ and mass density 
$\rho$ of the background stratification.

On top of the relaxed snapshot mentioned above, we added a potential magnetic nullpoint distribution 
as illustrated in panel~(b) of Figure~\ref{sup:01}. 
The potential configuration was calculated from a prescribed axisymmetric distribution at the bottom boundary 
containing a circular positive polarity (the {\it parasitic} polarity) embedded in a negative background.
This imposed configuration is such that (a) there is a nullpoint at $(x_{np}, y_{np}, z_{np}) = (16, 16, 8)$~Mm; 
(b) at large heights, the magnetic field becomes asymptotically vertical and uniform with $B_z \to -10$~G, thus 
mimicking a coronal hole structure; and (c) at $z=0$, the total positive flux is $\Phi^{+}=4.9\times10^{19}$~Mx, 
the maximum positive vertical field strength is $B_z=141.8$~G, the parasitic polarity covers a circular patch of 
radius $6.3$~Mm, and the intersection of the fan surface with the horizontal plane has a radius of $12.4$~Mm. 
%

\section{OBSERVATIONS}\label{app:observations}

\subsection{The Swedish 1-m Solar Telescope (SST) Data.} 
We have used observations from the SST \citep{Scharmer_etal:2003} obtained on 2022 July 01 from 08:04:57 to 
08:13:03 UT centered at a CBP located at heliocentric coordinates $(x,y)$ = (176\arcsec,135\arcsec). 
To cover the whole CBP and surroundings within the field-of-view (FOV) of the SST instrumentation, we have 
created a mosaic constructed from $3\times3$ telescope pointings. 
%
    The SST data set contains the following spectral scans from the 
    CRisp Imaging Spectropolarimeter \citep[CRISP,][]{Scharmer_etal:2008}: 
    the \FeI\ 6173 \AA\ line in spectropolarimetric mode sampled in 14 positions
    between $-320$~m\AA\ and $+680$~m\AA\ from the line core, and 
    \Halpha\ 6563 \AA\ in spectroscopic mode sampled in 31 positions ranging 
    from $-1500$~m\AA\ to $+1500$~m\AA\ from the line core. 
    Each of the CRISP scans of the mosaic has a FOV of $\sim58\arcsec\times58\arcsec$
    and a spatial sampling of 0\farcs057 per pixel. 
    The CRISP 
    data were processed using the SSTRED data 
    reduction pipeline \citep{de-la-Cruz-Rodriguez_etal:2015, Lofdahl_etal:2021},
    which includes Multi-Object Multi-Frame Blind Deconvolution
    \citep[MOMFBD,][]{Van-Noort_etal:2005} image restoration. 
    Line-of-sight (LOS) magnetograms were obtained by Milne-Eddington 
    inversions of the \FeI\ 6173 \AA\ Stokes profiles \citep{de-la-Cruz-Rodriguez:2019}. 

\subsection{The Solar Dynamics Observatory (SDO) Data.} 
The coronal response of the targeted CBP is obtained from the Atmospheric Imaging Assembly
\citep[AIA,][]{Lemen_etal:2012} onboard SDO \citep{Pesnell_etal:2012} around 08:08:52 UT: the time at which the center of 
the SST mosaic was taken.
The photospheric magnetic field is obtained from the Helioseismic and Magnetic Imager 
\citep[HMI,][]{Scherrer_etal:2012}.
Regarding the alignment, cross-alignment is carried out in all the AIA channels to the HMI continuum; then, the 
SDO images are manually aligned with the SST ones.

\section{FORWARD MODELING}\label{app:synthesis}

\subsection{EUV and UV Forward Modeling.}
For the forward modeling in the EUV and UV, the expressions of optically thin radiative transfer are used, 
assuming, additionally, coronal abundances \citep{Feldman:1992} and ionization equilibrium. 
The emissivity corresponding to any given transition between electron configuration levels $i$ and $j$ of a given 
ionic species of an atomic element (e.g., Fe IX, or Si IV) is given by
\begin{equation}
    \epsilon = n_H\ n_e\ G(T, n_e,\nu_{ij}) 
    \label{eq:emiss}
\end{equation}
where $n_e$ and $n_H$ are the electron and hydrogen number densities, respectively; $\nu_{ij}$ is the radiation 
frequency of the emitted photon; and $G(T, n_e,\nu_{ij})$ is the corresponding gain function (or contribution 
function), which already contains the abundance of the atomic element. 
The emitted intensity is calculated by integrating Equation~\ref{eq:emiss} along the LOS in the numerical box. 

For the SDO/AIA coronal synthetic images, we have used CHIANTI \citep{Del-Zanna_etal:2021} from the Solar Soft package,
calculating the gain functions for the atomic transitions in the relevant spectral range of the different AIA filters 
and weighing with the corresponding effective Area Function. 
The result is then transformed into SDO/AIA count numbers. 
To account for the possible obscuration effects from cool and dense features in the EUV images, we have included 
absorption by neutral hydrogen, neutral helium, and singly-ionized helium in the integration of the emissivity along 
the LOS adding an absorption factor $e^{-\tau}$ following the recipes and procedures in the literature 
\citep{De-Pontieu_etal:2009, Anzer_Heinzel:2005}. 

For the synthetic 174~\AA\ images of the Extreme Ultraviolet Imager of the High Resolution Imager (EUI-HRI)
\citep{Rochus_etal:2020}; on Solar Orbiter (SoLO) \citep{Muller_etal:2020}, we have employed a contribution 
function privately provided by Dr. Fr\'ed\'eric Auch\`ere, member of the Solar Orbiter team.

For the synthetic IRIS \citep{De-Pontieu_etal:2014} transition region counterpart, the contribution function of the 
Si IV 1393.755 \AA\ line is obtained through a simple procedure (Chianti Solar Soft routine {\tt
  ch\_synthetic.pro}  with the flag {\tt /goft}), multiplying the output by the silicon abundance relative to hydrogen. 
This procedure and the conversion to IRIS count numbers are detailed in a recent paper
\citep{Nobrega-Siverio_etal:2017}.

After obtaining the intensity images, we have degraded the results to match the spatial resolution of the 
instruments, namely, 1\farcs5  for SDO/AIA  \citep{Lemen_etal:2012}; 0\farcs33 for IRIS \citep{De-Pontieu_etal:2014}; 
and  $200$~km for EUI-HRI at perihelion \citep{Rochus_etal:2020}.

\subsection{X-ray Forward Modeling.}
The X-ray synthetic images have been computed to mimic observations taken by the X-ray telescope (XRT) 
\citep{Golub_etal:2007} onboard Hinode \citep{Kosugi_etal:2007}. 
In particular, we have combined the {\tt make\_xrt\_wave\_resp.pro} and {\tt make\_xrt\_temp\_resp.pro} routines 
from Solar Soft to obtain the spectral and temperature response for the Al-poly channel of the telescope. 
The result is correspondingly multiplied by the squared electron number density and then degraded to the
2\farcs0 instrument resolution \citep{Golub_etal:2007}.

\subsection{Chromospheric Forward Modeling.}
To obtain \Halpha\ synthetic images, we have used the 3D non-LTE radiative transfer code Multi3D 
\citep{Leenaarts_Carlsson:2009}. 
Multi3D evaluates the radiation field in full 3D taking the horizontal structure of 3D model atmospheres into account. 
To reduce the 3D radiative transfer computation cost, we have halved the horizontal resolution of the model atmosphere. 
We further optimized the depth scaling of the atmospheric quantities to improve the convergence time and to reduce 
the effect of an insufficiently sampled optical depth scale onto the synthesized \Halpha\ line profiles.  
\Halpha\ was synthesized from a five-level plus continuum hydrogen model atom assuming Doppler line profiles for the 
\Lyalpha\ and \Lybeta\ lines which affect the level populations of the \Halpha\ transition \citep{Leenaarts_etal:2012}.
We have reduced the spatial resolution of the synthetic images to the SST/CRISP instrumental values,
namely, 0."13 at 6301~\AA\ \citep{Scharmer_etal:2008}.

\bibliography{references}
\bibliographystyle{aasjournal}

\end{document}